\documentclass[printer]{aa}
\usepackage{graphicx}
\usepackage{txfonts}
\usepackage[colorlinks,citecolor=blue]{hyperref}
\def \arcsec {$^{''}$}

\graphicspath{{./}}

\begin{document}

 \title{Observations of pores and surrounding regions with CO 4.66 $\mu$m lines by BBSO/CYRA}

   \author{Yongliang Song\inst{1}
          \and Xianyong Bai\inst{1,2}
          \and Xu Yang\inst{3,4}
          \and Wenda Cao\inst{3,4}
          \and Han Uitenbroek\inst{5}
          \and Yuanyong Deng\inst{1,2}
          \and Xin Li\inst{1}
          \and Xiao Yang\inst{1}
          \and Mei Zhang\inst{1,2}}

   \institute{Key Laboratory of Solar Activity, National Astronomical Observatories, Chinese Academy of Sciences, Beijing 100101, China\\
             \email{ylsong@nao.cas.cn; xybai@nao.cas.cn}
         \and
             School of Astronomy and Space Science, University of Chinese Academy of Sciences, Beijing 100049, China
         \and
             Big Bear Solar Observatory, New Jersey Institute of Technology, Big Bear City, CA 92314-9672, USA\\
             \email{wenda.cao@njit.edu}
          \and
             Center for Solar-Terrestrial Research, New Jersey Institute of Technology, 323 Martin Luther King Boulevard, Newark, NJ 07102, USA
          \and
            National Solar Observatory, University of Colorado Boulder, 3665 Discovery Drive, Boulder, CO 80303, USA}

\abstract
     {Solar observations of carbon monoxide (CO) indicate the existence of lower-temperature gas in the lower solar chromosphere. We present an observation of pores, and quiet-Sun, and network magnetic field regions with CO 4.66 $\mu$m lines by the Cryogenic Infrared Spectrograph (CYRA) at \emph{Big Bear Solar Observatory.} }
   {We used  the strong CO lines at around 4.66 $\mu$m to  understand the properties of the thermal structures of lower solar atmosphere in different solar features with various magnetic field strengths. }
   {Different observations with different instruments were included:  CO 4.66 $\mu$m imaging spectroscopy by CYRA, Atmospheric Imaging Assembly (AIA) 1700 \AA\ images, Helioseismic and Magnetic Imager (HMI) continuum images, line-of-sight (LOS) magnetograms, and vector magnetograms.
   The data from 3D radiation magnetohydrodynamic (MHD) simulation with the  \emph{Bifrost} code are also employed for the first time to be compared with the observation.
   We used the Rybicki-Hummer (RH) code to synthesize the CO line profiles in the network regions.
   }
   {The CO 3-2 R14 line center intensity changes to be either enhanced or diminished with   increasing magnetic field strength, which should be caused by different heating effects in magnetic flux tubes with different sizes. We find several ``cold bubbles'' in the  CO 3-2 R14 line center intensity images, which can be classified into two types. One type is located in the quiet-Sun regions without magnetic fields. The other type, which has rarely been reported in the past, is near or surrounded by magnetic fields. Notably, some are   located at the edge of the magnetic network. The two kinds of cold bubbles and the relationship between cold bubble intensities and network magnetic field strength are both reproduced by the 3D MHD simulation with the \emph{Bifrost} and RH codes. The simulation also shows that there is a  cold plasma blob near the network magnetic fields, causing the observed cold bubbles seen in the  CO 3-2 R14 line center image.
   }
   {Our observation and simulation illustrate that the magnetic field plays a vital role in the generation of some CO cold bubbles.}


   \titlerunning{Observations of Pores and Surrounding regions with CO 4.66 $\mu$m Lines}
   \authorrunning{Song et al.}

   \maketitle

\section{Introduction}
\label{sec:intro}
The fundamental absorption lines of carbon monoxide (CO) at around 4.66 $\mu$m have been observed since the 1970s \citep[e.g.,][]{NoyesHall1972, Ayres1981, Ayres1986, Farmer1989, Solanki1994, Uitenbroek1994}.
The discovery of these strong absorption lines indicates the  existence of a large amount of cool gas with a temperature that can be as low as 3700 K in the solar atmosphere \citep[e.g.,][]{NoyesHall1972, AyresTesterman1981, AyresBrault1990}.
From observations and simulation models, it is found that the formation height for these lines is in a range from the upper photosphere to the middle chromosphere \citep[e.g.,][]{Solanki1994, Uitenbroek1994, Clark1995, Uitenbroek2000a, Ayres2002, Asensio2003, Wedemeyer2005,Stauffer2022}.
However, the temperature in the temperature minimum region from the classical model of solar atmosphere   is about 4400 K, based on the emissions of strong Mg II h and k lines and ultraviolet (UV) lines \citep[e.g.,][]{Staath1995, Uitenbroek1997, Carlsson1997}.
This provides a great challenge for us to understand the thermal structures of the lower solar atmosphere.

Early observations of sunspots by \citet{Uitenbroek1994} and \citet{Clark2004} with CO lines revealed the inverse Evershed flow in the penumbra; in other words, the flow moves from the quiet-Sun (QS) region into the umbra.
\citet{PennSchad2012} found highly sheared flow in the vicinity of sunspots after an X-class flare.
Observations of QS and sunspots both show oscillations in CO line center intensity  and Doppler velocity, with a period of 3 minutes and 5 minutes, respectively \citep[e.g.,][]{NoyesHall1972, Uitenbroek1994, Solanki1996,Uitenbroek2000a,Li2020}.

A two-component temperature model has been proposed that contains hot flux tubes and cool nonmagnetic atmospheres; the latter is  named CO-cooled clouds or COmosphere  \citep{Ayres1981, Ayres1986}.
However, observations show that the CO line intensity for the  QS is very dynamic, presumably driven by the granulation motions and the p-mode oscillations \citep{Uitenbroek1994, AyresRabin1996, Uitenbroek2000a}.
This indicates that no nonmagnetic areas can be constantly as cool as ``cool clouds''. Instead, cool episodes can be found everywhere in the QS regions \citep{Uitenbroek2000a}.
The result is reproduced in a 3D numerical simulation of the nonmagnetic lower solar atmosphere with the radiation chemo-hydrodynamics code CO5BOLD \citep{Wedemeyer2006}.
Recently,  the cool low chromospheric gas is also found in a QS region from Atacama Large Millimeter/submillimeter Array (ALMA) observations other than infrared CO lines \citep{Loukitcheva2019}.

In this paper we present the observations of pores, and quiet Sun and network magnetic field regions with CO 4.66 $\mu$m lines by the state-of-the-art instrument \emph{Cryogenic Infrared Spectrograph (CYRA)} of the \emph{Goode Solar Telescope (GST)} at Big Bear Solar Observatory (BBSO) \citep{Goode2012}.
We analyze the relationship between CO lines emission and magnetic field.
The data from radiation magnetohydrodynamic (MHD) simulation  with the  \emph{Bifrost} code are also employed in this work.
Section 2 presents the observations and data reduction.
The radiation MHD simulation is described in Section 3.
The results are presented in Section 4.
We give our conclusion and discussion in Section 5.

\section{Observations and data reduction}
\label{sec:obs}

\begin{figure*}
\centering
\includegraphics[trim=0.0cm 1.0cm 0.0cm 0.0cm,width=1.0\textwidth]{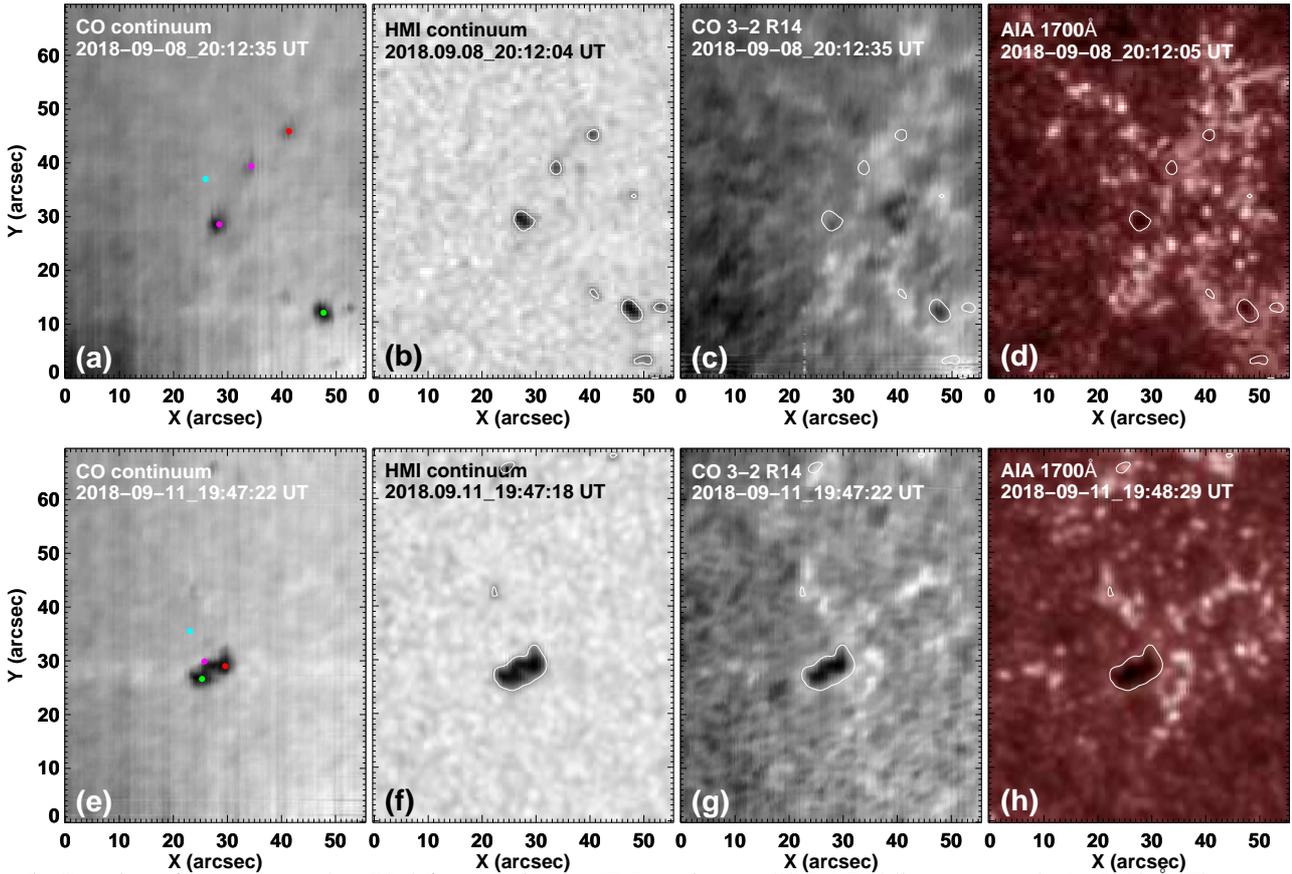}
\caption{
Overview of pores observed at CO 4.6 $\mu$m continuum, HMI continuum, CO 3-2 R14 line center, and AIA 1700 \AA.
The upper (a-d) and lower  (e-h) panels were observed on September 8 and 11, 2018, respectively.
The white contours indicate the enhanced absorption features (pores) in HMI continuum intensity images.
The colored dots in panels (a) and (e) give the positions of the CO spectral lines shown in Fig. \ref{fig:2}(a, b).
}
\label{fig:1}
\end{figure*}

CYRA is a novel instrument that probes the rich spectral regime of 1.0--5.0 $\mu$m in the infrared (IR), which  is the first fully cryogenic spectrograph in any solar observatory. It is designed to study the solar activities in the photosphere to chromosphere by observing the unexplored IR regime \citep{Cao2010, Cao2012, Yang2020}.
We observed two regions near the disk center that contain several pores and the surrounding network magnetic field regions with CYRA at $\sim$20:12 UT on September 8 and $\sim$19:47 UT on September 11, 2018, respectively.
We performed a 75-step raster scan over these two regions with the scanning step of 0.8\arcsec.
The field of view is about $\sim$68\arcsec $\times$ 55\arcsec (See Figure \ref{fig:1}) and the pixel resolution is about $\sim$0.148\arcsec.
We used the  high-quality calibrated data of CYRA, in which the dark and flat field corrections, removing residual bad pixels and hot pixels, and the non-uniform illumination of spectrum corrections have all been taken into account \citep{Yang2020}.

The data from the \emph{Atmospheric Imaging Assembly} \citep[AIA,][]{Lemen2012}  and the \emph{Helioseismic and Magnetic Imager } \citep[HMI,][]{Scherrer2012}  on board the \emph{Solar Dynamics Observatory} \citep[SDO,][]{Pesnell2012} are also employed in this research.
AIA observes the full-disk Sun in two ultraviolet (UV) and seven extreme-ultraviolet (EUV) passbands.
The spatial pixel size is about $\sim0.6^{\prime\prime}$ and the time cadences of the UV and EUV observations are 24 s and 12 s, respectively.
Here we only use the AIA 1700~\AA\ images.
HMI observes the full-disk photosphere at six wavelength positions across the spectral line of Fe {\footnotesize I}  6173~\AA .
HMI provides continuum images, Dopplergrams, line-of-sight (LOS) magnetograms, and vector magnetograms.
The time cadences of these data are 45 s, 45 s, 45 s, and 12 minutes, respectively, and 
 the spatial pixel resolution is about $\sim$0.5\arcsec.
Here we use HMI continuum images, LOS, and vector magnetograms.

\section{Radiation MHD simulation}
\label{sec:rmhd}

The 3D parallel numerical code \emph{Bifrost} is designed to simulate the  stellar atmosphere from the convection zone to the corona \citep{GudiksenNordlund2005, Gudiksen2011}.
The simulation considers various physical and boundary conditions by solving MHD partial differential equations with a Cartesian grid.
To solve these MHD equations the sixth-order differential operator, fifth-order interpolation procedures, and a third-order Hyman time-stepping scheme are used in the code \citep{Gudiksen2011}.
As a   highlight of \emph{Bifrost}, three modules are included to handle the radiative transfer:  optically thin radiative transfer, chromospheric radiation approximation, and full radiative transfer computations \citep{Gudiksen2011}.

In this work we employ a 3D MHD numerical simulation model of solar atmosphere for an enhanced network magnetic field region calculated with the   \emph{Bifrost} code (en024048\_hion, http://sdc.uio.no/search/simulations).
This simulation contains $\sim504\times504\times496$ grid points corresponding to a physical volume of  $\sim24\times24\times17~Mm^3$  extending from the  convective zone (about -2.5~Mm) to the corona (about 14.5~Mm).
The horizontal grid spacing is $\sim$48~km, while the vertical grid spacing varies from $\sim$19~km to $\sim$119~km.
The magnetic field strength in the photosphere is about $\sim$50~G on average.  And the distance between two dominant opposite polarity regions is $\sim$8~Mm.
Non-equilibrium hydrogen ionization is included in the simulation.
The data gives 157 time steps (T=385--541) for the evolution of the atmosphere.
For our purpose, we selected one time step (T=385) simulation result and calculated a part of the whole physical volume, $\sim8.62\times8.14\times1.7~Mm^3$ (see Figure \ref{fig:6}).
The height is from the convective zone top (about -0.6~Mm) to the middle chromosphere (about 1.1~Mm).
The parallel RH code is used to synthesize the CO line profiles \citep{Rybicki1991, Rybicki1992, Uitenbroek2001, Pereira2015}.

\section{Results}
\label{sec:res}
\subsection{Results from CYRA observation}

\begin{figure*}
\centering
\includegraphics[trim=0.0cm 0.6cm 0.0cm 0.0cm,width=1.0\textwidth]{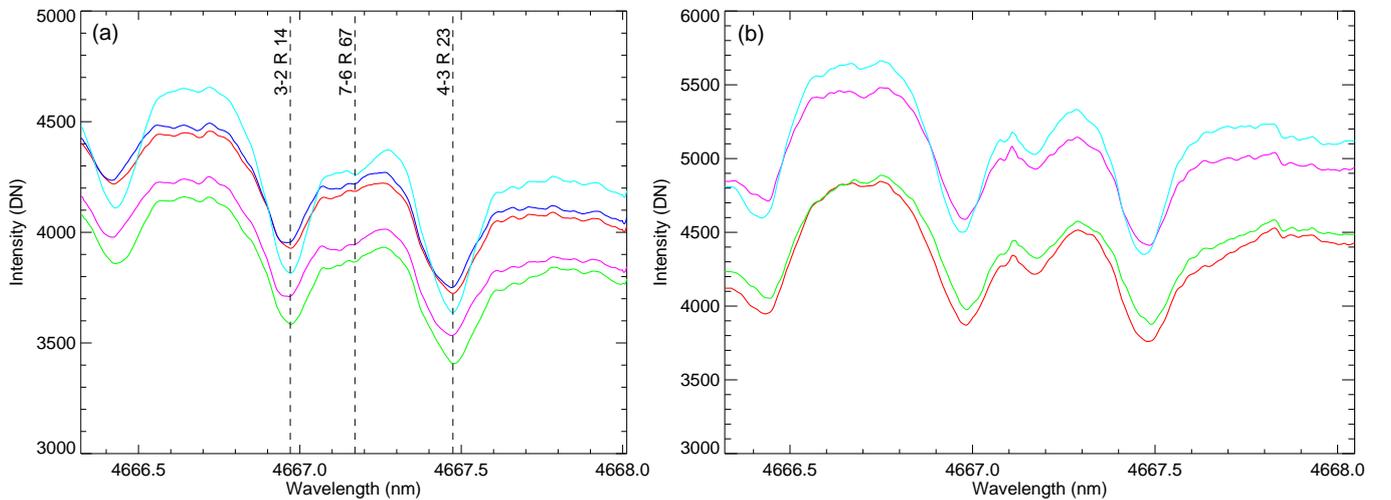}
\caption{
Plots of CO spectrum around 4.66 $\mu$m in different regions.
Panels (a) and (b) are for pores and the quiet Sun (cyan) indicated by the colored dots in Fig. \ref{fig:1}(a) and (e).
The dashed lines in panel (a) indicate three typical CO lines around 4.66 $\mu$m.
The left panel (a) and right panel (b) are for September 8 and 11, 2018, respectively.
}
\label{fig:2}
\end{figure*}

\begin{figure*}
\centering
\includegraphics[trim=0.0cm 0.8cm 0.0cm 0.0cm,width=1.0\textwidth]{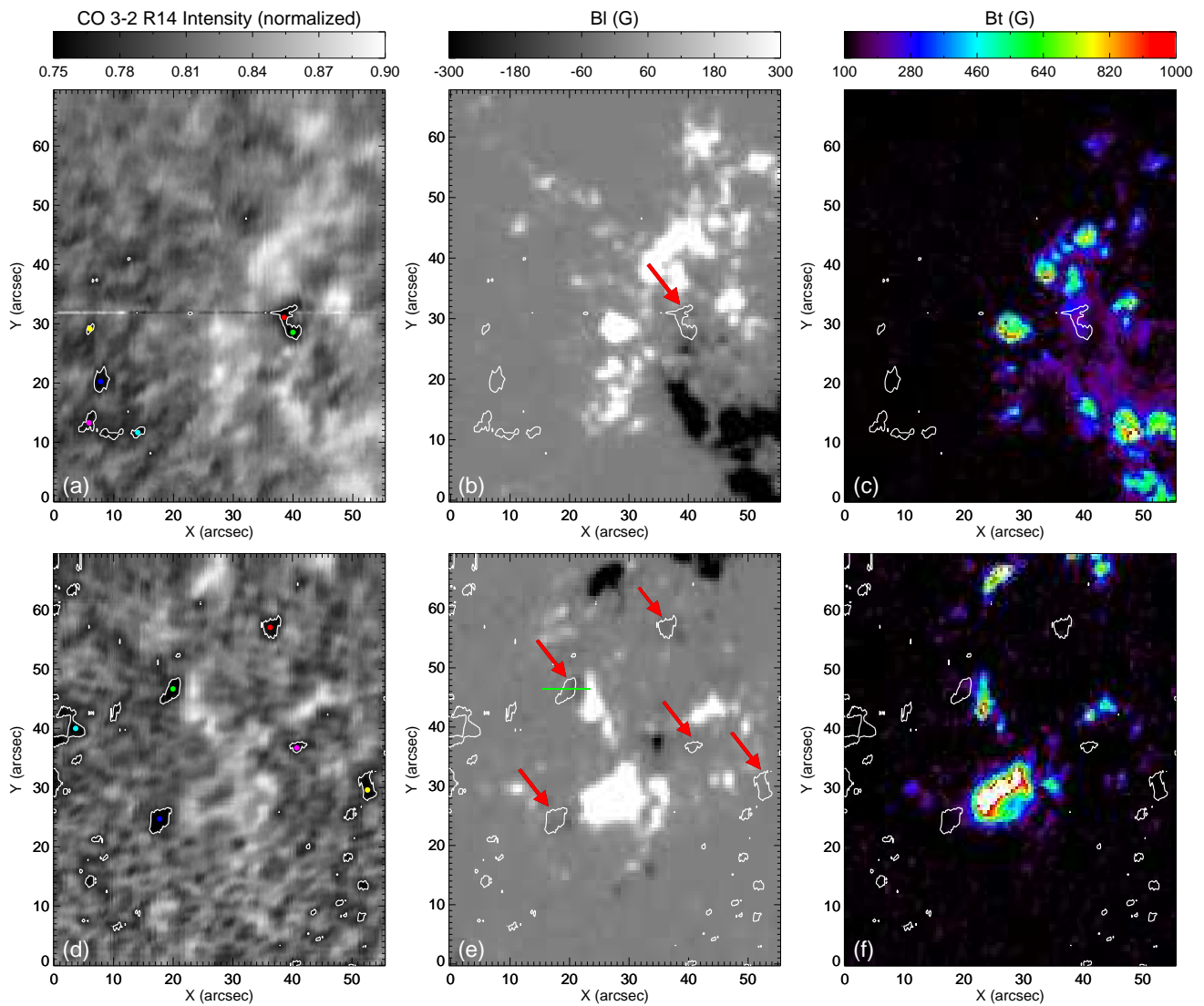}
\caption{
Normalized CO 3-2 R14 line center intensity images, HMI line-of-sight magnetograms, and horizontal magnetic field.
The top (a-c) and bottom  (d-f) panels are for September 8 and 11, 2018, respectively.
The white contours indicate  the dark regions\ (cold bubbles) in the normalized CO 3-2 R14 line center intensity map with a level of $I_{3-2 R14}/I_{continuum} =0.77$.
The red arrows indicate the cold bubbles near or surrounded by magnetic fields.
The colored dots in panels (a) and (d) give the positions where we show the normalized CO spectral lines in Fig. \ref{fig:4}(a, b).
The green line in panel (e) is the slit on which we show the changes of the CO 3-2 R14 line center intensity and $B_l$ strength in Fig. \ref{fig:8} (a).
}
\label{fig:3}
\end{figure*}

\begin{figure*}
\centering
\includegraphics[trim=0.0cm 0.6cm 0.0cm 0.0cm,width=1.0\textwidth]{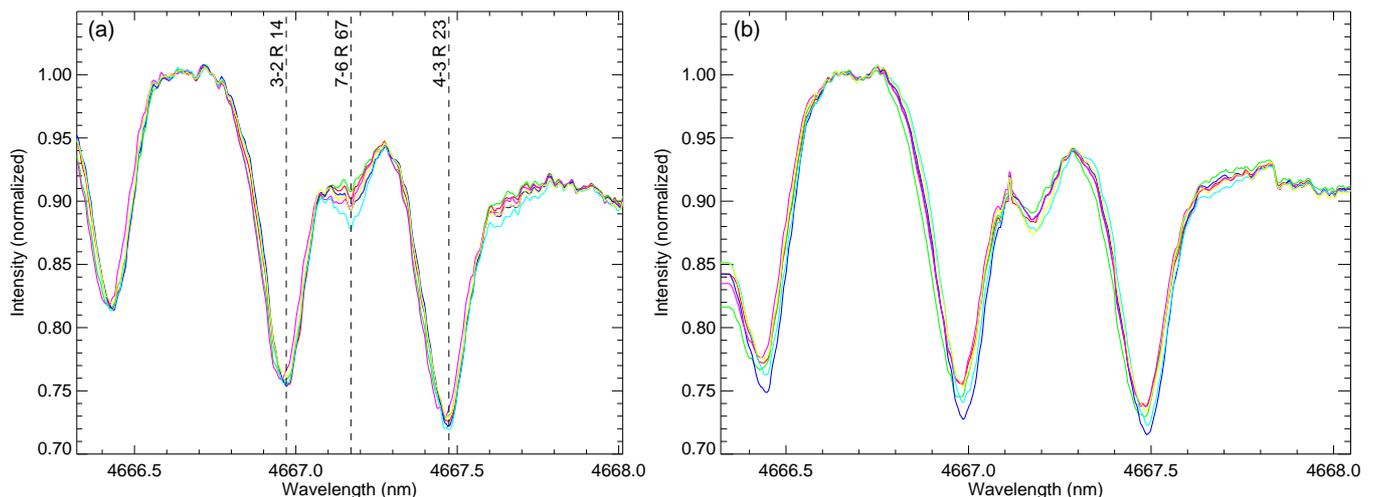}
\caption{
Normalized ($I/I_{continuum} $) CO spectral lines in cold bubbles indicated by colored dots in Fig.\ref{fig:3}(a) and (d).
The left panel (a) and right panel (b) are for September 8 and 11, 2018, respectively.
The dashed lines in panel (a) indicate three typical CO lines around 4.66 $\mu$m.
}
\label{fig:4}
\end{figure*}

\begin{figure*}
\centering
\includegraphics[trim=0.0cm 0.5cm 0.0cm 0.0cm,width=0.8\textwidth]{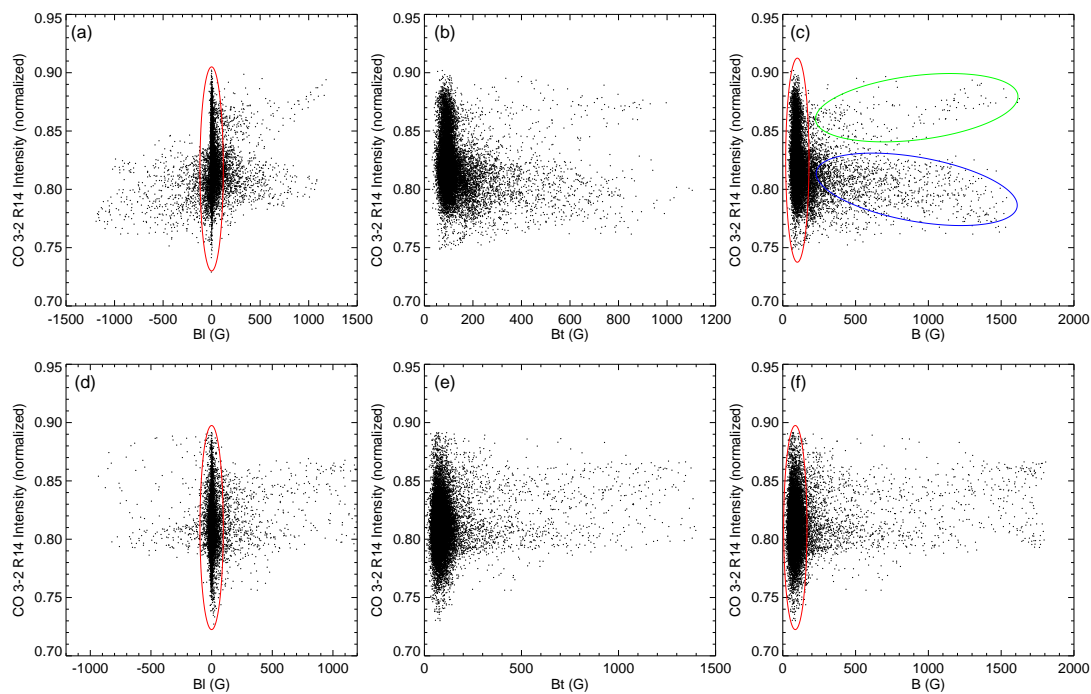}
\caption{
Scatter plots showing the relationships between the normalized CO 3-2 R14 line center intensity ($I_{3-2 R14}/I_{continuum} $) and the line-of-sight magnetic field ($B_l$), the horizontal magnetic field ($B_t$), and the total magnetic field strength ($B$).
The top panels (a-c) and bottom panels (d-f) are for September 8 and 11, 2018.
The red ellipses indicate the QS regions where the CO 3-2 R14 line center intensity changes in an extensive range.
The green ellipse in panel (c) shows that the CO 3-2 R14 line center intensity is brighter when the magnetic field strength is stronger. The blue ellipse reveals the opposite relationship, which means the CO 3-2 R14 line center intensity is inversely proportional to the magnetic field strength.
}
\label{fig:5}
\end{figure*}

\begin{figure*}
\centering
\includegraphics[trim=0.0cm 0.8cm 0.0cm 0.0cm,width=0.9\textwidth]{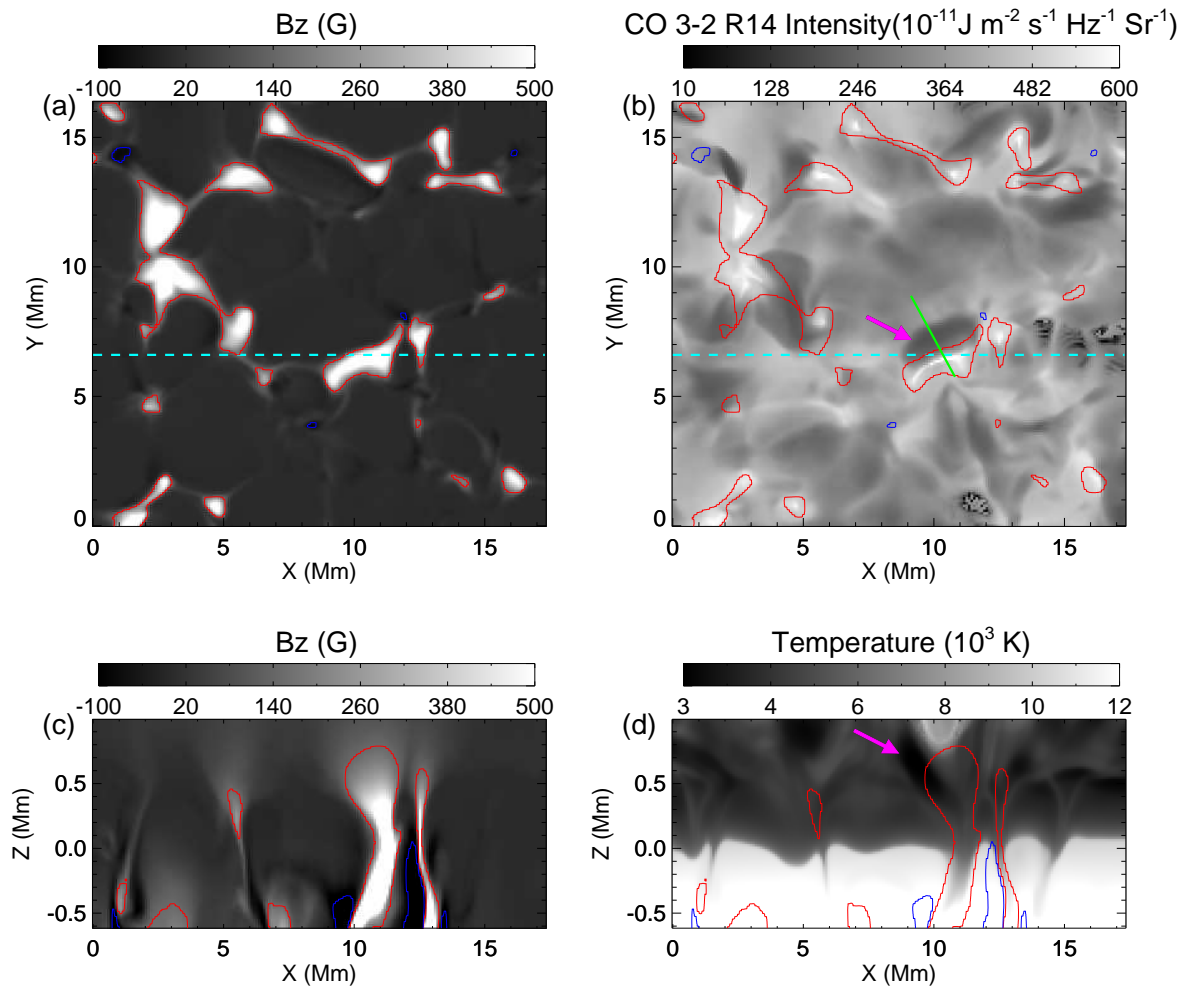}
\caption{
Radiation MHD simulation results of network magnetic field region with \emph{Bifrost} code.
Panels (a) and (b) show the vertical component of the magnetic field ($B_z$) and synthetic CO 3-2 R14 line center intensity map.
Panels (c) and (d) show the strength of $B_z$ and temperature on the slice indicated by the dashed cyan lines in panels (a) and (b).
The red and blue contours indicate the positive and negative magnetic fields with 150 G and -50 G, respectively.
The magenta arrow in panel (b) indicates a typical cold bubble from the CO 3-2 R14 line center intensity map, which is located near the edge of the magnetic network. Correspondingly,  the temperature of the bubble regions at the height of 500 km is   lower than the other regions (magenta arrow in   panel (d)).
The green line in panel (b) indicates the slit showing the relationship of CO 3-2 R14 line center intensity and $B_z$ strength arranged in Fig. \ref{fig:8} (b).
}
\label{fig:6}
\end{figure*}

\begin{figure*}
\centering
\includegraphics[trim=0.0cm 0.5cm 0.0cm 0.0cm,width=0.93\textwidth]{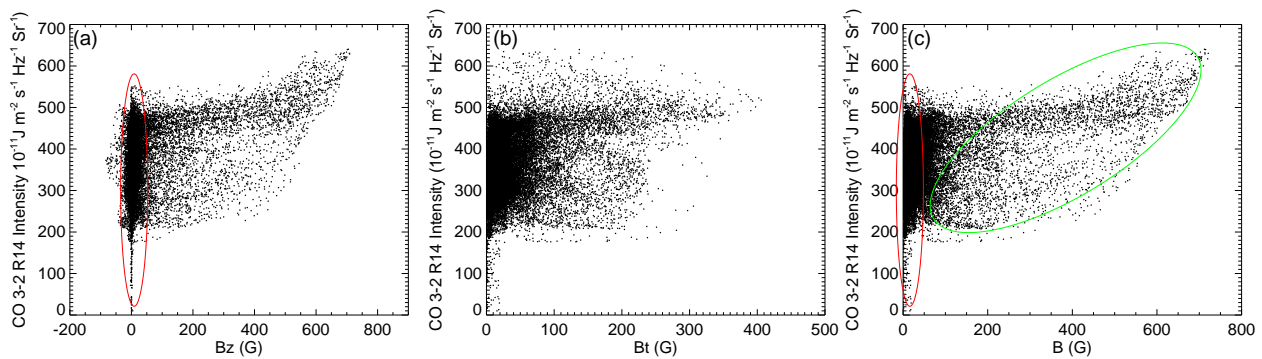}
\caption{
Similar to Fig.\ref{fig:5}, but for scatter plots showing the relationships between the CO 3-2 R14 line center intensity and the vertical magnetic field ($B_z$), the horizontal magnetic field ($B_t$), and the total magnetic field strength ($B$) for the  \emph{Bifrost} data seen in Fig.\ref{fig:6}.
}
\label{fig:7}
\end{figure*}

\begin{figure*}
\centering
\includegraphics[trim=0.0cm 0.3cm 0.0cm 0.0cm,width=0.99\textwidth]{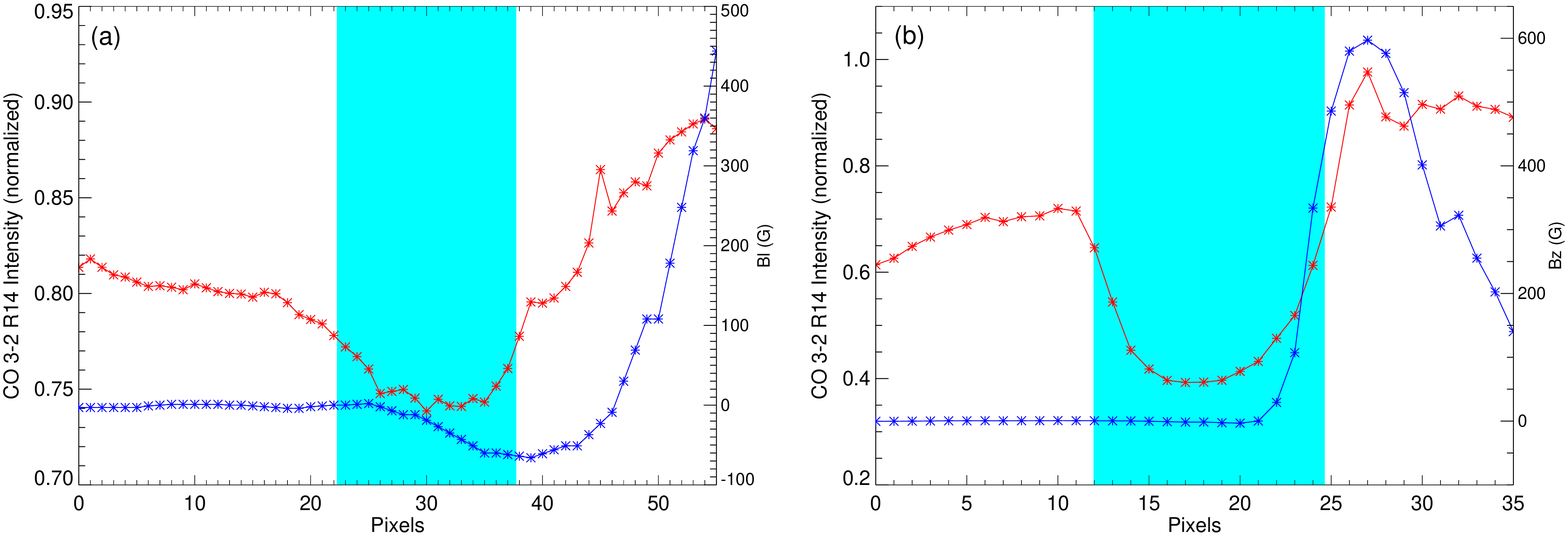}
\caption{
Normalized CO 3-2 R14 line center intensity (red) and $B_l$ or $B_z$ strength (blue).
Panel (a) is along the slit indicated by the green line in Fig. \ref{fig:3}(e), and panel (b) is along the slit shown in Fig. \ref{fig:6}(b).
The cyan shading indicates the edge of the network magnetic field regions.
}
\label{fig:8}
\end{figure*}

Figure \ref{fig:1} shows the pores and surrounding regions on September 8 and 11, 2018, observed at the CO continuum, CO 3-2 R14 line center, HMI continuum, and AIA 1700~\AA.
The pores observed on September 8 are tiny, with an average diameter of less than $\sim$5\arcsec.
Only one relatively large pore, with a size of
$\sim$10\arcsec, was observed on September 11.
CO continuum and HMI continuum images seem to be similar.
The IR granulations are also recognizable in the CO images, while  clearer structures are found in HMI continuum images.
The difference is a combined result caused by different wavelengths (IR 4660 nm and optical 617.3 nm) with varying formation heights and under different observing conditions (ground-based and space-borne).

Comparing the CO 3-2 R14 line center intensity maps with AIA 1700 \AA\ images, they all present bright network structures.
AIA 1700 \AA\ emission is mainly from the  temperature minimum region and photosphere  \citep{Lemen2012}, while the  CO 3-2 R14 line formation height is from the upper photosphere to lower chromosphere.
Similar bright network structures seen in both channels are due to the similar formation height.
However, CO 3-2 R14 line center intensity images show some dark patches that are not significant in the AIA 1700~\AA\ images.
Since CO molecule lines have a source function close to local thermodynamic equilibrium (LTE) in the formation region, their intensities reveal the temperatures in the formation atmosphere  \citep{AyresWiedemann1989, Uitenbroek2000a}.
Here, we named these dark patches ``cold bubbles,''  indicating these regions owing to low-temperature gas or plasma.
These cold bubbles are contoured in Figure \ref{fig:3}.

We compare the spectral lines near 4.66 $\mu$m in the pores and the QS region   in Figure \ref{fig:2}(a) and \ref{fig:2}(b) for the two days, respectively.
Four points in the four larger pores are selected for September 8; for September 11 we selected three points in the pore.
The locations of the selected points are indicated by  colored dots in Figure \ref{fig:1} (a) and \ref{fig:1}(e).
The shapes of these spectral profiles of different pores or at different pore positions are similar, but their intensities can be   different.
This may be caused by different temperatures in the pores, due to different heating effects in differently size magnetic flux tubes \citep{Spruit1981}.
The spectral profiles in the QS region (cyan line), on the contrary, are notably different.
Its continuum emission is more significant than that in pores. However, at CO absorption line centers such as 3-2 R14, its emission can be lower than some pores.
This should result from different temperature structure models in pores and a QS atmosphere.
In particular, the pores are regions with strong magnetic fields, while the quiet Sun is not.

Figure \ref{fig:3} shows images for normalized CO 3-2 R14 line center intensity ($I_{3-2 R14}/I_{continuum}$), LOS magnetic field ($B_l$), and horizontal magnetic field ($B_t$).
We indicate some cold bubbles (dark patches) by white contours with a level of $I_{3-2 R14}/I_{continuum}=0.77$.
The shapes are not uniform; the sizes vary from less than $\sim$1\arcsec to about $\sim$5\arcsec$\times$6\arcsec.
These cold bubbles can be classified into two types according to their locations.
One type is located in the QS region without any magnetic fields.
The other type is near or surrounded by magnetic fields, as shown by the  red arrows in Figure \ref{fig:3}(b) and \ref{fig:3}(e).
The cold bubble indicated by the red arrow in panel (b) observed on September 8 is surrounded by the network magnetic fields.
The strength of $B_l$ in this region is weaker than the surrounding area, 
but the horizontal magnetic field $B_t$ is   stronger, with a strength of about 150$\sim$300~G.
The cold bubbles in panel (e) indicated by red arrows observed on September 11 are mainly near or on the edge of the magnetic network.
The horizontal magnetic field $B_t$ in these regions is weaker than the  cold bubble   surrounded by the network magnetic field in panel (b).
This result suggests  that cold bubbles  exist with different magnetic field structures, including unipolar, bipolar, or even more complicated structures.
In a recent work, \citet{Stauffer2022} also reported the CO  cold bubbles  observed in the region surrounding a large pore, indicating a potential role of the magnetic field in the formation of these   cold bubbles.

Normalized spectral lines ($I/I_{continuum}$) of different  cold bubbles  (colored dots in Figure \ref{fig:3}(a) and \ref{fig:3}(d)) are shown in panels (a) and (b) of Figure \ref{fig:4}.
Most of them are very similar, especially for the  cold bubbles  observed on September 8 (panel  \ref{fig:3}(a) and \ref{fig:4}(a)).
For the observation on September 11, the spectral line at the position indicated by a blue dot (Fig.\ref{fig:3} (d)) in the  cold bubble  near the pore shows a   greater depth in the absorption line center (Fig.\ref{fig:4} (b)).
The results suggest that most  cold bubbles  share a similar formation temperature except for specific regions.

Figure \ref{fig:5} shows the scatter plots between normalized CO 3-2 R14 center intensity ($I_{3-2 R14}/I_{continuum} $) and the LOS magnetic field $B_l$, the horizontal magnetic field $B_t$, and the total magnetic field strength $B$.
The CO 3-2 R14 line intensity in the QS region within the red ellipses changes significantly from the largest absorption depth to the smallest absorption depth, indicating the QS atmosphere is very dynamic. The temperature in this region is not constant as previous studies have shown \citep[e.g.,][]{Uitenbroek1994, AyresRabin1996, Uitenbroek2000a}.

As the magnetic field becomes much stronger, the CO 3-2 R14 line center intensity can be brighter or darker, showing a positive (green ellipse in Fig. \ref{fig:5}(c)) or negative correlation (blue ellipse in Fig. \ref{fig:5}(c)).
This phenomenon should be caused by different thermal effects on the solar atmosphere for magnetic flux tubes with varying sizes in the formation of pores or network fields \citep{Spruit1981}.
However, these two trends are not apparent in the region observed on September 11 (see  panel (f)).
The possible reason is that only one big pore was observed on September 11, and the strongest magnetic fields are located in the pore (see Figs. \ref{fig:1},  \ref{fig:3}(e), and \ref{fig:3}(f)).
Thus, we do not see   changes in CO 3-2 R14 line center intensity with different sizes of magnetic flux tubes.

\subsection{Results from the Bifrost simulation}

There are very few CO 4.6 $\mu$m observations at the pore or network magnetic field regions in the past. Moreover, the 3D MHD simulations from \citet{Wedemeyer2006} and \citet{Uitenbroek2000a} only include the   nonmagnetic solar photosphere and low chromosphere and do not have a network magnetic solar atmosphere. The state-of-the-art MHD model from the  \emph{Bifrost}  code contains an enhanced network, giving us an unprecedented opportunity to check whether the synthetic CO emission map under such a condition is consistent with that observed from CYRA.

Figure \ref{fig:6} presents the result of the  radiation MHD simulations with \emph{Bifrost} code.
Comparing the network magnetic field (Fig. \ref{fig:6}(a)) and synthetic CO 3-2 R14 line center intensity map with RH code (Fig. \ref{fig:6}(b)), we find a significant part of  the cold bubbles  with lower line center intensities located on or near the edge of network magnetic field beside the QS areas. The simulation result is consistent with our observations, as shown in Figure \ref{fig:3} and corresponding descriptions.
For example, the   cold bubble  indicated by a magenta arrow in panel (b) is   located on the edge of a network magnetic field region.
We take  cross sections of the vertical magnetic field and temperature for this  cold bubble  from the position given  by the dashed cyan lines in Figure \ref{fig:6}(a) and \ref{fig:6}(b).
The sliced maps are displayed in the bottom panels of Figure \ref{fig:6}.
The magenta arrow in panel (d) indicates a low-temperature plasma blob.
Its temperature is lower than 4000~K, and it is situated in height from $\sim$400~km (upper photosphere) to $\sim$900~km (lower chromosphere), which is in the typical formation height of CO 3-2 R14 line. 
The  ``cold plasma blob'' located on the edge of the network magnetic field region produced the observed  cold bubble  in the CO line center map.

Figure \ref{fig:7} presents the scatter plots between the  CO 3-2 R14 center intensity and magnetic field $B_z$, $B_t$, $B_t$ for the  \emph{Bifrost} data.
Similarly to Figure \ref{fig:5}, we see the CO 3-2 R14 line intensity in the QS region changes within a great range, and the intensity becomes much brighter with the stronger magnetic field.
It is also consistent with our observations, as shown in Figure  \ref{fig:5} and corresponding descriptions.

Figure \ref{fig:8} exhibits the plots of CO 3-2 R14 line center intensity and magnetic field strength ($B_l$, $B_z$) along the slits in Figure \ref{fig:3}(e) and Figure \ref{fig:6}(b), respectively.
Both observation and simulation share a similar change.
From the outer part of the magnetic network to the inner part, the CO 3-2 R14 line center intensity first changes to darker, then changes to brighter.
The darkest part is located at the edge of the magnetic network.
Observations and numerical simulations both show that the cold bubbles  can exist in an environment with magnetic fields.

\section{Conclusion and discussion}
\label{sec:dis}
We carried out the observations of pores and surrounding network magnetic field regions in the IR CO 4.66 $\mu$m with the new generation and fully cryogenic BBSO/CYRA. 
To better understand the observations, we also synthesized the CO line center intensity map of a network magnetic field region for the first time with RH code from the \emph{Bifrost}-based 3D radiation MHD model. 
In the paper, we mainly focus on the relationship between CO lines emission and magnetic field, which was rarely investigated in previous studies \citep[e.g.,][]{Ayres1981, Ayres1986, Uitenbroek1994, Ayres1998, Wedemeyer2006}. 
Supporting the previous studies, significant intensity changes of the CO 3-2 R14 line are observed in the QS region without magnetic fields, indicating the solar atmosphere in this region is highly dynamic in temperature. 
As the magnetic field gets stronger, we find the CO 3-2 R14 line center intensity will change to be brighter or darker, which results from different thermal effects by flux tubes with different sizes \citep{Spruit1981}.

We also find some  cold bubbles  from CO 3-2 R14 line center intensity images.
They can be classified into two types based on their locations, located in the quiet Sun and near or on the edge of the magnetic network. Our simulation successfully reproduced these two types of  cold bubbles.
The second type (cold bubbles    with magnetic fields) is rarely reported, due to the limited observations. Our study shows this type should be prevalent from the two days of  observations and the simulation.

Two types of  cold bubbles  indicate different cooling mechanisms in the solar atmosphere.
One is without magnetic fields in the quiet Sun, as previous studies show \citep[e.g.,][]{Uitenbroek2000a, Wedemeyer2006}, and the other includes magnetic fields. The deepest part of the  CO 3-2 R14 line center intensity of some  cold bubbles    corresponds to the edge of the magnetic network (Fig. \ref{fig:8}).
According to the MHD simulation, the magnetic field can prevent heating from the lower solar atmosphere and form a  cold plasma blob. Observations and simulations both illustrate that the magnetic field plays a vital role in the generation and dynamic evolution of some CO  cold bubbles.

The question arises of how to understand the mechanism of the magnetic field in producing CO  cold bubbles.
In   early observations, \citet{Ayres1981} reported ``thermal shadows'' around small sunspots.
The author explained that a thermal shadow is formed by the near-surface horizontal magnetic field, which suppresses convection and yields cooler temperature in the upper atmosphere.
It is reasonable, although it  may need strong and low horizontal magnetic fields.
In a recent work by \citet{Stauffer2022}, they found several CO  cold bubbles  that surround a large pore.
By calculating the radially averaged three-minute power surrounding the pore in different observational lines, they proposed that the canopy magnetic field prevents the acoustic energy from propagating upward to heat the lower chromosphere.
Although there are some differences between the pictures of \citet{Ayres1981} and \citet{Stauffer2022}, they both  emphasize the important role of the horizontal magnetic field in some  cold bubbles.
Similarly,  cold bubbles  near the pore are also recorded in our two days of  observations.
The very existence of a strong horizontal magnetic field in a region of  cold bubbles  (Fig.\ref{fig:3} (b) and (c)) supports the  explanations of both  \citet{Ayres1981} and \citet{Stauffer2022}.

However, based on our observation and MHD simulation, we also propose another possible mechanism:   massive CO molecules generated by cooling and overshooting of granulation adiabatic expansion horizontally accumulate to the boundary of supergranulation.
 The boundary of supergranulation is where the network magnetic fields is located. 
Masses of CO molecules can produce radiative cooling and form deep-absorption CO lines \citep{AyresRabin1996, Uitenbroek2000a, Ayres2002, Penn2014}.
It should be noted that  the cold bubbles  reported by \citet{Ayres1981} and \citet{Stauffer2022} are mainly near or surround small sunspots or pores.
Unprecedentedly, we notice some  cold bubbles  located near or just on the edge of the magnetic network both from observations and MHD simulations (Figs. \ref{fig:3}(e), \ref{fig:3}(f) and Figs. \ref{fig:6}(b), \ref{fig:6}(d)).
Particularly, the  cold bubbles  near the network magnetic field present even stronger absorptions than those inside the network region (Fig.\ref{fig:6}(b)).
The deepest part of CO 3-2 R14 line center intensity  corresponds to the edge of the magnetic network (Fig.\ref{fig:8}).

Our results illustrate the vital role of the magnetic field in the generation of some CO  cold bubbles, and highlight the importance of CO observations to better understand the thermal structures of solar atmosphere near the temperature minimum region of different solar features. As  CYRA is taking an increasing amount of raster scan data, and the newly constructed  Cryogenic Near Infra-Red Spectro-Polarimeter (Cryo-NIRSP) from the Daniel K. Inouye Solar Telescope (DKIST) will begin to perform  observations, we are looking forward to analyzing the dynamic evolution of  cold bubbles  in QS and magnetic field regions. Combing CO observations and the evolutions of 3D MHD simulation having magnetic fields, for example the  \emph{Bifros}t and \emph{MURaM} codes, we will be able to estimate the role of   magnetic fields in their generation and evolution in detail.

\begin{acknowledgements}
The authors acknowledge the use of data from the Goode Solar Telescope (GST) of the Big Bear Solar Observatory (BBSO) and Solar Dynamics Observatory (SDO). This work is supported by National Key R\&D Program of China No. 2021YFA1600500 and NSFC grants 12173049, 11803002, 11873062, 11973056, 12003051. BBSO operation is supported by US NSF-1821294 and New Jersey Institute of Technology. GST operation is partly supported by the Korea Astronomy and Space Science Institute and the Seoul National University. X. Yang and W. Cao acknowledge support from US NSF grants - AGS-1821294 and AST-2108235.

\end{acknowledgements}

\bibliographystyle{aa}
\bibliography{bibliography}

\end{document}